\documentclass[12pt,preprint]{aastex}            

\newcommand{\ergcm}{ergs~cm$^{-2}$~s$^{-1}~$}

\newcommand{\mdot}{$\dot{{m}}$ } 
\def\ni{\noindent}  
\def\about{$\sim$} 

\def\erg/cm2sec{ergs~cm$^{-2}$~s$^{-1}$}   
\def\ergcm2{ergs~cm$^{-2}$}   
\def\cm2{cm$^2$} 
\def\la{\hbox{\rlap{\raise.3ex\hbox{$<$}}\lower.8ex\hbox{$\sim$}\ }}
\def\ga{\hbox{\rlap{\raise.3ex\hbox{$>$}}\lower.8ex\hbox{$\sim$}\ }}

\def\deg{$^{\circ}$~}

\received{}
\accepted{}
\begin{document}

\title{EXITE2 Observation of the SIGMA Source GRS 1227+025} 

\author{J.E. Grindlay}
\affil{Harvard-Smithsonian Center for Astrophysics, Cambridge, MA 02138}

\author{Y. Chou}
\affil{National Tsing Hua University, Hsinchu, Taiwan, Republic of China }

\author{P.F. Bloser}
\affil{NASA/Goddard Space Flight Center, Greenbelt, MD 20771 }


\and
\author{T. Narita}
\affil{College of the Holy Cross, Worcester, MA 01610}

\email{jgrindlay@cfa.harvard.edu}

\begin{abstract}

We report the EXITE2 hard X-ray imaging of the sky around
3C273.  A 2h observation on May 8, 1997, shows a $\sim$260 mCrab source
detected at $\sim$4$\sigma$ in each of two bands (50-70 and 70-93 keV) 
and located $\sim$30' from 3C273 and consistent in
position with the SIGMA source GRS1227+025. The EXITE2 spectrum 
is consistent with a power law with photon index 3 and large low 
energy absorption, as indicated by the GRANAT/SIGMA results. No source was
detected in more sensitive followup EXITE2 observations in 2000 and
2001 with 3$\sigma$ upper limits of 190 and 65 mCrab, respectively.
Comparison with the flux detected by SIGMA shows the source to be
highly variable, suggesting it may be non-thermal and beamed and thus
the first example of a ``type 2'' (absorbed) Blazar.  Alternatively it
might be (an unprecedented) very highly absorbed binary system 
undergoing accretion disk instability outbursts, possibly either
a magnetic CV, or a black hole X-ray nova.

\end{abstract}

\section{Introduction}

GRS1227+025 was discovered serendipitously with SIGMA
\citep{Bassani91, Jourdain92} in the 40-80 keV energy band during an
observation of the QSO 3C273.  Separated from 3C273 by only $\sim$15',
GRS1227+025 (hereafter GRS1227) was detected at 5.5$\sigma$ whereas
3C273 was not detected.  The source has a significantly steeper
spectrum (photon index 3$^{+1.3}_{-0.9}$, \cite{Bassani91})
compared to 3C273, which was detected in subsequent observations, with
its usual flatter photon index of $\sim$1.5.  Upper limits (15-30 keV)
with ART-P (simultaneous with SIGMA) showed that the spectrum must be
heavily absorbed: N$_H \sim$1.5$\times$10$^{25}$ cm$^{-2}$ for a power
law spectrum with photon index 3 \citep{Bassani93}.  Followup observations
done with SIGMA and OSSE (which could only measure the combined flux
of 3C273 and GRS1227) showed that GRS1227 is probably variable by a
factor 2-5 on $\sim$1 day time scales \citep{Bassani93, Johnson95}.
No obvious soft X-ray counterpart was found in the SIGMA 
error circle (5 arcmin, 1$\sigma$) with ROSAT (e.g. \cite{Leach96}),
although an Einstein source (1E 1227+0224) in the SIGMA error 
circle (1$\sigma$) was identified with a QSO at z=0.57 that would 
have to be of ``type 2'' (heavily self-absorbed) if it were indeed 
the counterpart \citep{Grindlay93}.

\section{EXITE2 Observation and Data Analysis}

The second-generation Energetic X-ray Imaging Telescope Experiment
(EXITE2) was a phoswich (NaI(Tl)/CsI(Na)) imaging telescope
designed to detect and image cosmic X-ray sources in the broad hard
X-ray energy band (20 - 600 keV) from a high-altitude balloon.  The
details of the payload are described in \cite{Lum94, Manandhar95, 
Chou98, Chou01, Chou03}.  EXITE2 flew three
times (from Ft. Sumner, New Mexico) in 1997, 2000 and 2001 for
scientific observations, which included the Crab Nebula, the X-ray
binaries Cyg X-1 and 4U 0614+09, the microquasar
GRS 1915+105, the Seyfert galaxy NGC 4151 and 
QSO 3C273/GRS1227.  The data analysis methods and
the observation results of the Crab and Cyg X-1 are reported in
separate papers \citep{Bloser02a, Chou03}.  Other results
from the 2000 and 2001 flights, including separate experiments 
for the development of imaging Cd-Zn-Te detectors (Bloser et 
al 2002b, Jenkins et al 2003) are reported elsewhere.

\begin{figure}
\plotone{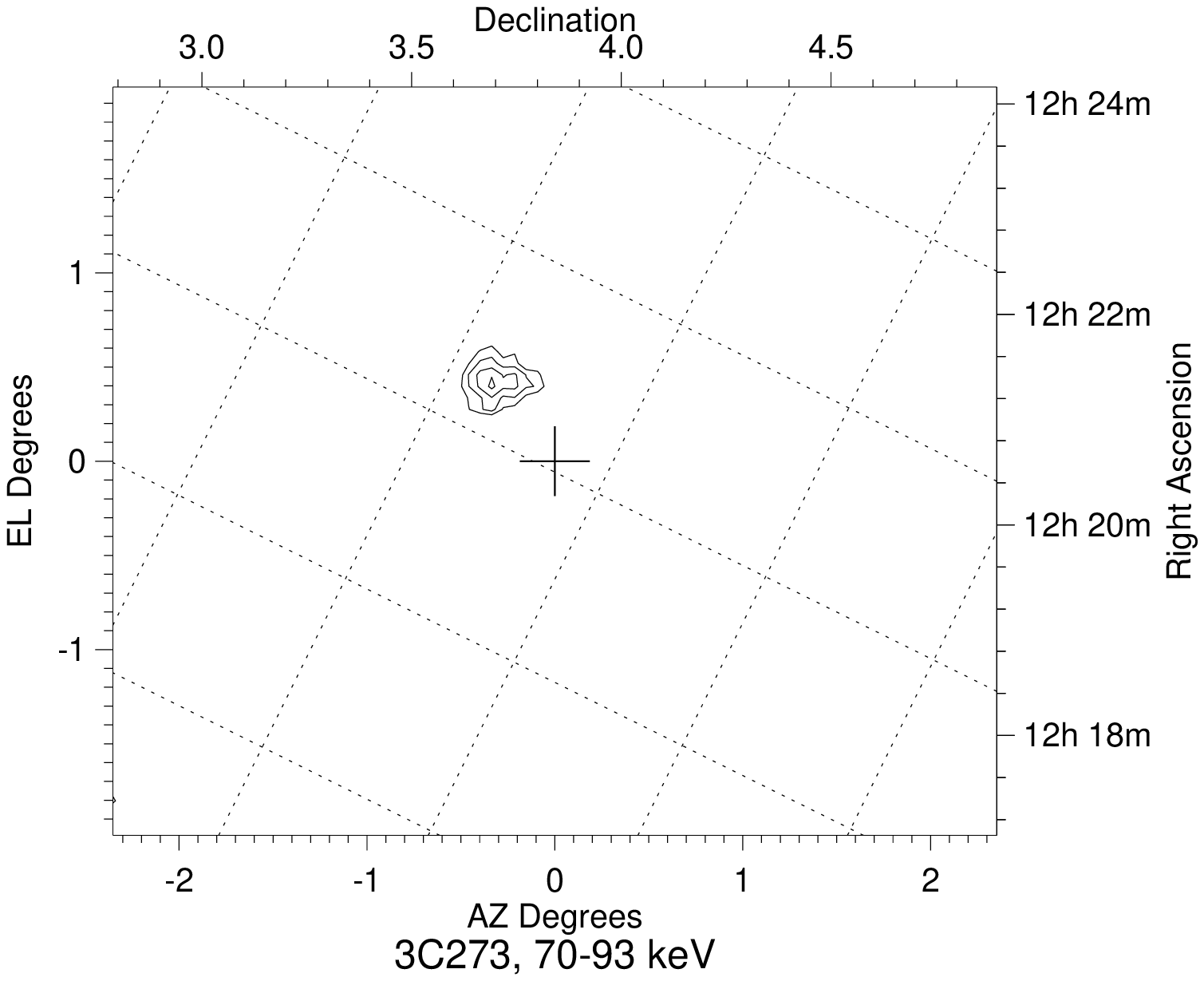}
\caption{Image of GRS1227+025 obtained with EXITE2 on UTMay 8.17, 1997, 
in the 70 - 93 keV band. Contours begin at 2.5$\sigma$ and 
increase in steps of 0.5$\sigma$ to $\sim4\sigma$. 
The source is independently detected at $\sim4\sigma$  
in the 50-73 keV band. The expected position of 3C273 is 
shown by the cross; the source detected is at a position (J2000) 
compatible with the SIGMA detection of GRS1227+025. \label{3c273img}}
\end{figure}

For the 1997 flight, during UT03:20 to UT05:32 on May 8 (local time 21:20
to 23:32 on May 7), the telescope was pointing to within 5 arcmin of
the QSO 3C273 at (night) flight altitude $\sim$115,000 feet.  The data were
analyzed by the EXITE2 standard analysis methods described in 
Chou (2001), Bloser et al. (2002a) and Chou et al. (2003).  
The sky images showed that no significant 
($>$3$\sigma$) detection of  3C273 was obtained during the $\sim$2 hour 
observation.  However, a source $\sim$30' away from 3C273 was detected
in the EXITE2 energy band 5 (50-70 keV) and band 6 
(70-93 keV) at 3.9$\sigma$ and
4.1$\sigma$, respectively.  We estimated aspect errors as \la 5' from
agreement of the Crab and Cyg X-1 images with their source positions.  
The image for band 6 is shown as Figure~\ref{3c273img}.  

The source detected by EXITE2 in 1997 near 3C273  was located
at $\alpha$(2000)= 12$^h$31$^m$13$^s$, $\delta$(2000)= +2$^\circ$18',
with positional uncertainty $\pm \, \sim$7 arcmin
(1$\sigma$). Although 
$\sim$18' away from the GRS1227 location reported by Jourdain et 
al (1992) and plotted in 
Figure~\ref{grspos}, the 2$\sigma$ error circles for SIGMA vs. EXITE2 
overlap. The detected energy bands and the combined 5$\sigma$
significance of the EXITE2 detection are both comparable to the 
original SIGMA detection, further supporting the reality of a 
highly variable source at this location and distinct from 3C273.  
The fluxes measured during the 2 hours observation were
(6.6$\pm$1.7)$\times$10$^{-4}$ photons cm$^{-2}$ s$^{-1}$ keV$^{-1}$ and
(3.5$\pm$0.85)$\times$10$^{-4}$ photons cm$^{-2}$ s$^{-1}$
keV$^{-1}$, for band 5 and 6 respectively, a factor
of $\sim$10 brighter than SIGMA 1990 observations.
The combined band 5-6 flux was (4.94$\pm$0.91)$\times$10$^{-4}$ photons
cm$^{-2}$ s$^{-1}$ keV$^{-1}$ ($\sim$260 mCrab).  No source was
significantly detected in band 4 (37-50 keV) or band 7 (93-127 keV)
with 3$\sigma$ upper limits 1.8$\times$10$^{-3}$ photons cm$^{-2}$
s$^{-1}$ keV$^{-1}$ and 1.2$\times$10$^{-4}$ respectively.
We found that a spectrum of power law index $\sim$3 as reported by
Bassani et al. 1991 for GRS1227 is consistent with 
the EXITE2 1997 observation results (see Figure~\ref{grsspec}).  
Alternatively, the source could still
have a flatter spectrum (e.g. photon index $\sim$1.7 as for typical
AGN) with high energy cutoff at $\sim$100 keV.

Additional observations of the field containing 3C273/GRS1227 
were made in the EXITE2 flights
on September 19 2000 (UT20:45 to UT21:30) and May 23 2001 (UT03:09 to
UT06:17 on May 24). No source was detected during the $\sim$45 minute
observation in the 2000
flight.  The 3$\sigma$ upper limit for the combined bands 5-6 (50 to
93 keV) was 3.6$\times$10$^{-4}$ photons cm$^{-2}$ s$^{-1}$
keV$^{-1}$ ($\sim$190 mCrab), which is a factor 1.5 below the 1997
value.  For the 2001 observation, the telescope was pointed to the
combined position of SIGMA/EXITE2, the intersection of 
the 2$\sigma$ error circles 
for EXITE2 and SIGMA 
at RA=12$^h$30$^m$14$^s$, DEC= +2$^\circ$14'53'' (J2000), 
as shown in Figure~\ref{grspos}.  
No source was detected during a $\sim$3
hour observation.  The 3$\sigma$ upper limit
for the combined bands 5-6 (50 to 93 keV) was 1.25$\times$10$^{-4}$
photons cm$^{-2}$ s$^{-1}$ keV$^{-1}$, or $\sim$65 mCrab, which 
is comparable to the original SIGMA detection.  The upper
limits for both EXITE2 observations are also plotted in Figure~\ref{grsspec}.
The combined EXITE2-SIGMA observations indicate that the source is
highly variable (by a factor of $\sim$10) at hard X-ray energies.

\section{Discussion}

The GRS1227 source presents a major puzzle.  Both EXITE2 and
SIGMA detect a source at approximately the same location 
(within 2$\sigma$) and with similar spectral properties which 
are very distinct from 3C273. Although the EXITE2 detection is 
marginal, as was the original SIGMA detection, the combined results 
are more likely to be real. GRS1227 must be highly 
self-absorbed and yet highly
variable. The SIGMA variations are a factor of $\sim$2 and
SIGMA-EXITE2 variation is nearly a factor of $\sim$20.  This is a
combination not seen before, and suggests a new class of object: a
type 2 (i.e. heavily absorbed) Blazar, since Blazars (BL Lacs) are
usually the only AGN which show such 
extreme variability.  However, it might also be
a Narrow Line Seyfert 1 (NLSy) galaxy (e.g. RX J2217.9--5941, see
\cite{Grupe01}) but must still be highly self-absorbed.  It may 
be similar to the NLSys with significant internal absorption 
and significant variability described by \cite{Fabian99}.
Alternatively it might be (an unprecedented) very highly absorbed
binary system, possibly a low luminosity magnetic CV undergoing
outburst, since intermediate polars at high inclination can also show 
significant internal absorption. 

\begin{figure}
\plotone{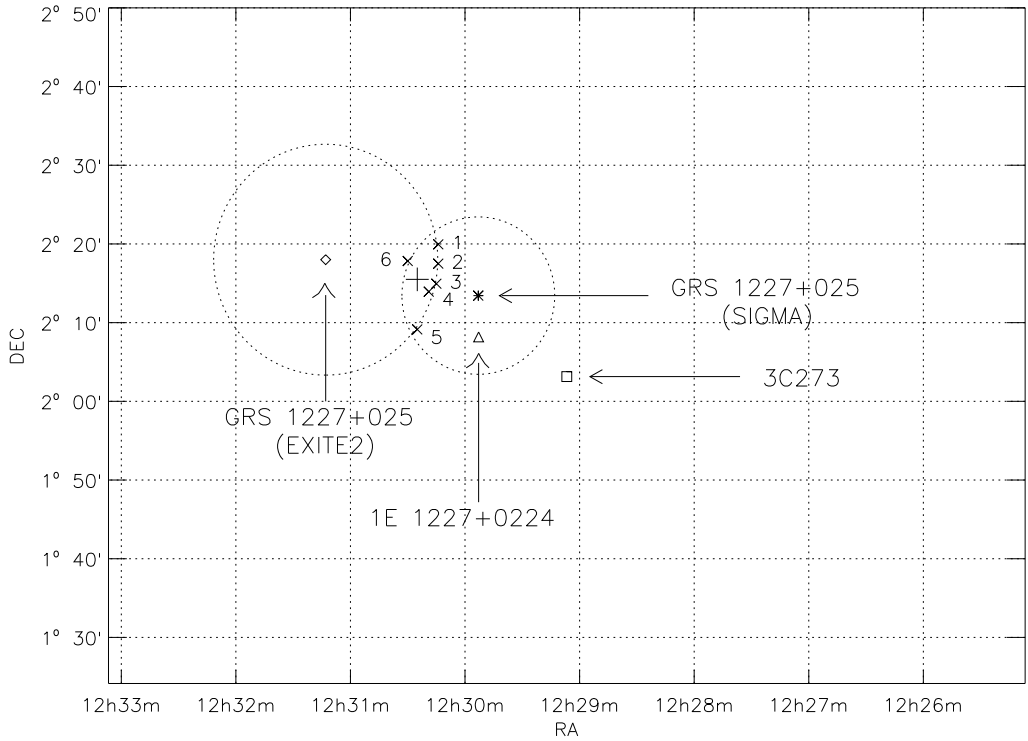}
\caption{GRS1227+025 source positions (plotted on J2000 grid) measured
(2$\sigma$ radius error circles) by 
SIGMA ($\sim$10') and by EXITE2 ($\sim$14'). 
The positions of
3C273 and 1E 1227+0224 are also shown.  The cross marked inside the
intersection of the 2$\sigma$ error circles is combined position of
EXITE2 and SIGMA.  The numbered source positions
(X) are galaxies (1,3,5,6) and radio galaxies (2,4) found in a
SIMBAD and NED search of the intersecting error circle for 
EXITE2/SIGMA. 
\label{grspos}}
\end{figure}

\begin{figure}
\plotone{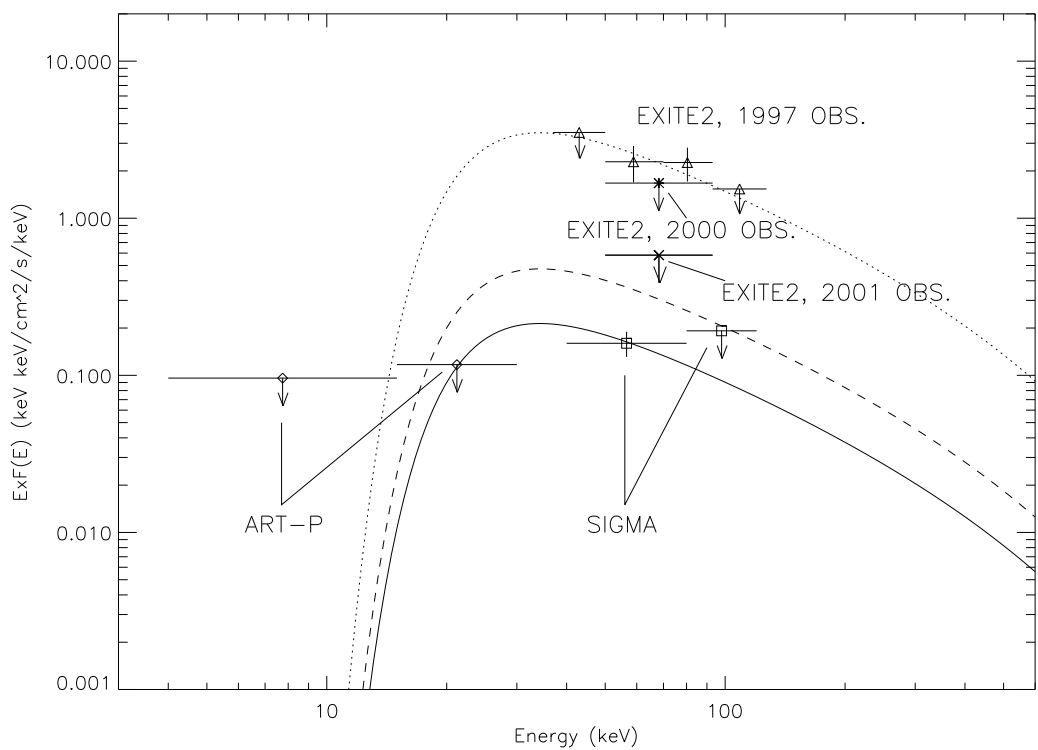}
\caption{Spectrum measured by the 1997 EXITE2 
observation (triangle data points and dotted curve), and the upper
limit from the 2000
observation (asterisk data point) and 2001 observation (``X'' data
point) vs. that measured in the SIGMA/ART-P observations 
(square and diamond points; from Jourdain et al. (1992) and dashed
curve, from Bassani et al. (1991)).  Each of the three curves is for a 
power law with photon index 3 and hydrogen column density
N$_H$=1.5$\times$10$^{25}$cm$^{-2}$ to fit the SIGMA/ART-P observation
results. \label{grsspec}}
\end{figure}

From the EXITE2 vs. SIGMA source error circles (2$\sigma$) shown in
Figure~\ref{grspos}, the combined source error box is $\sim$15'$\times$6',
centered on the EXITE2-SIGMA combined position RA=12$^h$30$^m$14$^s
(\pm$4'), DEC= +2$^\circ$14'53''($\pm$4') (J2000). At this
relatively uncrowded high latitude position, even a $\sim$2' position
could enable identification with an optical or radio variable.  
However the need for a \la1'
position is obvious: if this is, as suspected, a highly absorbed
non-thermal source (BL Lac), it may not be at all conspicuous in the
optical and appear as a simple elliptical galaxy.  The radio galaxies
numbered (2) and (4) in Figure~\ref{grspos}, or [SRK80]122803+023451
and 87GB[BWE91]1227+0230, are thus possible candidates.  However, in
the event that the radio emission is confined to the core, GRS1227
may not even be bright at radio wavelengths since it is likely
Compton-thick.  Given the space density of galaxies cataloged (SIMBAD
and NED) within the combined EXITE2/SIGMA error circles
(cf. Figure~\ref{grspos}), a \la1' source position is obviously needed
for a unique galaxy identification.

Source variability measurements can partly constrain the
underlying source nature. The generalized condition for 
maximum variability of an isotropically emitting source powered 
by accretion at rate \mdot and radiating luminosity L$_x$ is \\

$\Delta$L$_x$  \la  2 $\times$ 10$^{41}$ $\eta_{0.1}$ $\Delta$t ergs/s, \\

\ni
where $\eta_{0.1}$ = L$_x$/\mdot c$^2$ \about 0.1 is the 
likely maximum efficiency factor 
(cf. \cite{Guilbert83}). This can impose direct constraints 
on the source luminosity (and thus distance and nature) for an 
observed luminosity variation $\Delta$L$_x$ over time interval 
$\Delta$t. Therefore, a factor of 2 
increase on \about1d timescales (already suspected from SIGMA) 
would imply the source is beamed and thus a Type 2 Blazar 
if it were identified with 
an AGN at z \ga 0.1 (like 3C273) with L$_x$ \ga10$^{45}$ erg/s.

GRS1227 exhibits an unusual spectral shape.  The simultaneous 
detection by SIGMA (\ga40 keV) and upper limit at \la15 keV by 
the ART-P X-ray imager on GRANAT (Bassani 1991) 
indicated that the source is
highly self-absorbed (N${_H} \sim$10$^{24-25}$ cm$^{-2}$) and thus a
Compton-thick absorber (see \cite{Matt99} and
references therein).  If the spectrum above $\sim$40-50 keV is indeed
a power law with photon index $\sim$3, it is consistent with a 
Blazar or NLSy. Although NLSys are also highly variable, 
these have not been reported 
as Type 2 objects. As mentioned above, the EXITE2 spectral 
shape (Figure 3) is equally consistent with a flatter spectrum with photon
index 1.7 but high energy cutoff at $\sim$100 keV, as expected for 
a highly absorbed Seyfert 2 or Type 2 QSO. However the remarkable 
variability required argues against the Seyfert or Type 2 QSO 
interpretation and in favor of a Blazar, with highly variable 
non-thermal emission. In this case, large amplitude optical 
variability might also be expected, since \cite{Liller75} found a $\sim$5
magnitude brightening for the BL Lac PKS 1510-089 on the Harvard 
plates. Given the offset from 3C273, it is possible that significant 
optical variables could have been missed, and a search using 
the Harvard plates is in progress.

A magnetic CV interpretation is also possible even if the spectrum is
indeed a (moderately steep) power law: the magnetic CV AE Aqr can be
fit with a similar power law spectrum \citep{Eracleous96}, and
has been reported as a high energy gamma-ray source 
(cf. Meintjes and de Jager 2000 and references therein).  A magnetic CV
origin for GRS1227 might also explain the extreme flare variations:
the object could be a dwarf nova (like GK Per) undergoing outbursts.  
Although the X-ray emission in classical dwarf novae outbursts is 
usually soft, an outburst in a magnetic CV may contain significant 
non-thermal emission. Historical optical measurements of the 
could again provide an important test, since
the optical flux (from the disk) would be expected to 
increase by $\sim$5 mag.  If the EXITE2 1997 detection
was the peak of an outburst like that of GK Per, the expected X-ray
optical flux ratio F$_x$/F$_v$$\sim$10 would imply an outburst
magnitude V$\sim$10, and thus quiescence magnitude V$\sim$15-16, which 
is probably too bright to have been missed as a blue object in the Palomar
Green survey (Green, Schmidt and Liebert 1986), which revealed 
no such objects near 3C273.

A high latitude black hole X-ray binary, an X-ray 
nova (XN) like XTE J1118+480. which showed evidence for 
magnetic flares and non-thermal optical emission (Merloni, 
DiMatteo and Fabian 2000) is
conceivable.  Here the power law hard X-ray 
spectrum is natural; this is seen in
the very high state of BH novae (e.g. Nova Muscae, \cite{Esin97}) 
and is expected for the jet emission probable in a system like 
XTE J1118+480.  It
too would have to be highly self-absorbed, but - 
as with the CV hypothesis - is plausible if the system were at
inclination $\sim$90$^\circ$, and self-absorbed at the required column
density of N$_H\sim$10$^{25}$ cm$^{-2}$. Such a highly absorbed XN 
would be missed by the ASM on RXTE, and indeed no evidence for 
an ASM source is found at the time of the 1997 EXITE2 
observation (or the 2000 and 2001 EXITE2 
observations; cf. Figure 4). However, an optical 
outburst is again expected and should again be searched on historical 
monitoring (e.g. plate) data.  Thus the galactic 
binary/transient scenarios might be distinguished from the 
Type 2 BLazar or NLSeyfert interpretation by historical optical or 
radio variations, which are probably larger in the binary 
case given the relatively low luminosity secondary star expected.  
The lack of any (large amplitude) optical variable 
would support the Type 2 Blazar or NLSeyfert interpretation.

Definitive confirmation of the source existence, 
and of course a refined hard X-ray position and 
spectrum, are obviously needed.  Long after this paper 
was initially submitted, we have carried out an 
observation with INTEGRAL (IBIS telescope) to attempt to 
further measure the source position, spectrum and variability. 
Data have just been received and will be reported separately. 
An initial INTEGRAL/IBIS observation of 3C273 has been 
reported recently by Courvoisier et al (2003) which also 
provided a sensitive new upper limit ($\sim$10 mCrab) for 
emission from the position of GRS1227 -- a factor of $\sim$6   
below the SIGMA  detection and 2001 EXITE2 upper limit 
and $\sim$25 below the 1997 EXITE2 detection. Thus the source 
variability at \ga40 keV must be at least this factor.  
Wide-field XMM
observations of the 3C273 field 
should also be examined for evidence for a highly variable, highly 
absorbed (\la 10 keV) X-ray source which would enable identification.

\acknowledgments The authors thank T Gauron, J. Gomes, V. Kuosmanen,
F. Licata, G. Nystrom, A. Roy, R. Scovel (SAO) and J. Apple (MSFC) for
for the EXITE2 detector and gondola development, and NSBF personnel
for their excellent support of the balloon launches and flight
operations.  This work was partially supported by NASA grants
NAGW-5103 and NAG5-5279. PFB is a National Research Council Research
Associate at NASA/Goddard Space Flight Center. TN acknowledges support
from the College of the Holy Cross.

\clearpage

\clearpage

\clearpage 

\clearpage 
\begin{figure}
\plotone{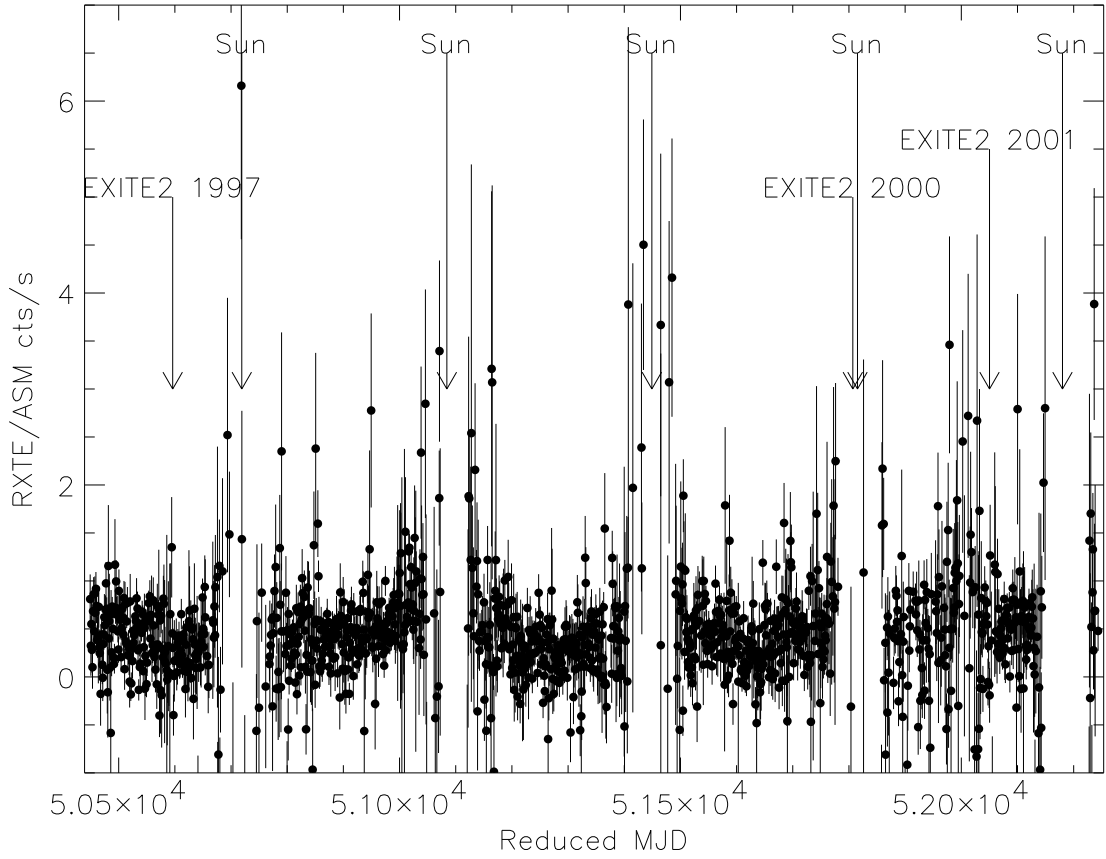}
\caption{RXTE/ASM light curve for 3C273 showing times of EXITE2 
observations of the field and 1997 detection vs. 2000 and 2001 
upper limits of GRS1227+025. The increased ASM count rate due 
to solar background is evident for the times (marked) when the 
Sun is closest ($\sim$4\deg) to 3C273 or GRS1227+025.}
\end{figure}

\end{document}